\documentstyle[amssymb,preprint,aps]{revtex}

\begin{document}
\title{Spectroscopy of $B_{c}$ meson in a semi-relativistic quark model using the
shifted large-N expansion method}
\author{Sameer M. Ikhdair\thanks{%
sameer@neu.edu.tr} and \ Ramazan Sever\thanks{%
sever@metu.edu.tr}}
\address{$^{\ast }$Department of Electrical and Electronic Engineering, Near East\\
University, Nicosia, North Cyprus, Mersin 10, via Turkey.\\
$^{\dagger }$Department of Physics, Middle East Technical University,\\
Ankara, Turkey.\\
PACS NUMBER(S): 11.10.St, 12.39.Pn, 14.40.-n}
\date{\today}
\maketitle

\begin{abstract}
We calculate the $c\overline{b}$ mass spectrum, the splitting values and
some other properties in the framework of the semi-relativistic equation by
applying the shifted large-N expansion technique. We use seven different
central potentials together with an improved QCD-motivated interquark
potentials calculated to two loops in the modified minimal-subtraction $%
\left( \overline{MS}\right) $ scheme. The parameters of these potentials are
fitted to generate the semi-relativistic bound states of $\ c\overline{b}$
quarkonium system in close conformity with the experimental and the present
available calculated center-of-gravity (c.o.g.) data. Calculations of the
energy bound states are carried out up to third order. Our results are in
excellent fit with the results of the other works.
\end{abstract}

\begin{center}
$\bigskip $
\end{center}

\begin{verbatim}
 
\end{verbatim}

\section{INTRODUCTION}

\noindent The spectroscopy of the $c\overline{b}$ system have already been
widely studied before in the framework of the heavy quarkonium theory [1].
After discovery of the $B_{c}$ was reported in 1998 by the (CDF)\
collaboration [2], the observed mass $M_{B_{c}}=6.40_{\pm 0.13}^{\pm 0.39}~%
{\rm GeV}~$has inspired new theoretical interest [1,3-9]. Bound state masses
have been estimated for the $B_{c}$ system which consists of heavy quarks
[1,3-9]. Hence, it can be reliably described by the use of the methods
developed for the $c\overline{c}$ and the $b\overline{b}$ spectra.

Quite recently, the revised analysis of the $B_{c}$ spectroscopy has been
performed in the framework of the potential approach [1,4,7,9] and QCD sum
rule [3,6]$.$ Kwong and Rosner [7] predicted the masses of the lowest $c%
\overline{b}$ vector (triplet) and pseudoscalar (singlet) states using an
empirical mass formula and a logarithmic potential. Eichten and Quigg [1]
gave a more comprehensive account of the energies and properties of the $c%
\overline{b}$ system that was based on the QCD-motivated potential of
Buchm\"{u}ller and Tye [10]$.$ Gershtein {\it et al.} [8] also published a
detailed account of the energies and decays of the $c\overline{b}$ system
using a QCD sum-rule calculation. Baldicchi and Prosperi have computed the $c%
\overline{b}$ spectrum based on an effective mass operator with full
relativistic kinematics [6]$.$ Moreover, they have fitted the entire
quarkonium spectrum. Fulcher {\it et al. }[4] extended the treatment of the
spin-dependent potentials to the full radiative one-loop level and thus
included effects of the running coupling constant in these potentials. He
also used the renormalization scheme developed by Gupta and Radford [11]$.$
Very recently, we have applied the shifted large-N expansion technique
(SLNET) [12-14] to study the $c\overline{b}~$ system in the context of
Schr\"{o}dinger equation [9].

In this work, we choose a group of seven central potentials [1,13-24] with a
strong coupling constant $\alpha _{s}$ to fit the spectroscopy of the
present spin-averaged energies of the existing quarkonium systems. We insist
upon strict flavor-independence of their parameters. We also extend this
study to an improved QCD-motivated interquark potential calculated to two
loops in the modified minimal-subtraction $\left( \overline{MS}\right) $
scheme [25,26]$.$ Such a potential contains an expansion to $\alpha _{s}$ to
at least 2-loop order and use of nontrivial methods in the interpolation
between long-distance (string constant) and short-distance behaviors [10]$.$

Since one would expect the average values of the momentum transfer in the
various quark-antiquark states to be different, some variation in the values
of the strong coupling constant and the normalization scale in the
spin-dependent should be expected.

In order to minimize the role of flavor-dependence, we use the same values
for the coupling constant and the normalization scale for each of the levels
in a given system and require that these values be consistent with a
universal QCD scale.

The contents of this article will be as follow: In section II, we present
the solution of the spinless Salpeter equation using the SLNET for the $c%
\overline{b}$ Salpeter spin-averaged binding energies. Section III presents
the proposed phenomenological central and the QCD-motivated potentials.
Summary and conclusions are presented in section IV. \ 

\section{WAVE EQUATION}

\noindent The relativistic wave Salpeter equation [14] is constructed by
considering the kinetic energies of the constituents and the interaction
potential. The spinless Salpeter (SS) for the case of two particles with
unequal masses $m_{q}$ and $m_{Q},$ interacting via a spherically symmetric
potential $V(r)$ in the center-of-momentum system of the two particles is
given by

\begin{equation}
\left[ 
\mathrel{\mathop{\sum }\limits_{i=q,Q}}%
\sqrt{-\Delta _{N}+m_{i}^{2}}+V(r)-M(q\overline{Q})\right] \psi \left( {\bf r%
}\right) =0,
\end{equation}
where the kinetic terms involving the operation $\sqrt{-\Delta _{N}+m_{i}^{2}%
}$ are nonlocal operators and $\psi ({\bf r})=Y_{\ell ,m}(\theta ,\phi
)R_{n,l}({\bf r})$ denotes the Salpeter's wave function. For heavy quarks,
the kinetic energy operators in Eq. (1) can be approximated, (cf. e.g.,
Jaczko and Durand of Ref. [27]), as

\begin{equation}
\mathrel{\mathop{\sum }\limits_{i=q,Q}}%
\sqrt{-\Delta +m_{i}^{2}}=m_{q}+m_{Q}-\frac{\Delta _{N}}{2\mu }-\frac{\Delta
_{N}^{2}}{8\eta ^{3}}+\;\cdots ,
\end{equation}
where $\mu =\frac{m_{q}m_{Q}}{m_{q}+m_{Q}}$ denotes the reduced mass and $%
\eta =\left( \frac{m_{q}m_{Q}}{m_{q}m_{Q}-3\mu ^{2}}\right) ^{1/3}.$\ $\mu $
is a useful mass parameter. This SS-type equation retains its relativistic
kinematics and is suitable for describing the spin-averaged spectrum of two
bound quarks of masses $m_{q}$ and $m_{Q}$ and total binding mass $M(q%
\overline{Q}).$ The Hamiltonian containing the relativistic corrections up
to order $(v^{2}/c^{2})$ is called as the generalized Breit-Fermi
Hamiltonian (cf. e.g., Lucha {\it et al. }of Ref.{\it \ }[21]). Therefore,
the spinless Salpeter equation can be written (in units $\hbar =c=1)$ [14] 
\begin{equation}
\left\{ -\frac{\Delta _{N}}{2\mu }-\frac{\Delta _{N}^{2}}{8\eta ^{3}}%
+V(r)\right\} R_{n,l}({\bf r})=E_{n,l}R_{n,l}({\bf r}),
\end{equation}
where $E_{n,l}=M_{n,l}(q\overline{Q})-m_{q}-m_{Q}$ refers to the Salpeter
quark binding energy with $M_{n,l}(q\overline{Q})$ is the
semirelativistic-bound-state masses of the non-self-conjugate atomlike meson
such as $c\overline{b}$ meson and $\Delta _{N}=\nabla _{N}^{2}.$\footnote{%
This approximation is correct to $O(v^{2}/c^{2}).$ The $\Delta _{N}^{2}$
term in (3) should be properly treated as a perturbation by using trial
wavefunctions [28].} Furthermore, in order to obtain a Schr\"{o}dinger-like
equation, the perturbed term in Eq. (3) is treated using the reduced
Schr\"{o}dinger equation [29]

\begin{equation}
p^{4}=4\mu ^{2}\left[ E_{n,l}-V(r)\right] ^{2},
\end{equation}
with $p^{4}=\Delta _{N}^{2},$ and consequently one would reduce Eq. (3) to
the Schr\"{o}dinger-type form [14]

\begin{equation}
\left\{ -\frac{\Delta _{N}}{2\mu }-\frac{\mu ^{2}}{2\eta ^{3}}\left[
E_{n,l}^{2}+V^{2}(r)-2E_{n,l}V(r)\right] +V(r)\right\} R_{n,l}({\bf r}%
)=E_{n,l}R_{n,l}({\bf r}).
\end{equation}
Further, the $N-$dimensional space operator in the spherical polar
coordinates is \ 
\begin{equation}
\Delta _{N}=\frac{\partial ^{2}}{\partial r^{2}}+\frac{N-1}{r}\frac{\partial 
}{\partial r}-\frac{L_{N-1}^{2}}{r^{2}},
\end{equation}
with $L_{N-1}^{2}=l\left( l+N-2\right) .$ After employing the following
transformation\ \ \ 
\begin{equation}
R_{n,l}(r)=\frac{u_{n,l}(r)}{r^{\left( N-1\right) /2}},
\end{equation}
and proposing the shifting parameter $\overline{k}=k-a$ with $k=N+2l,$ one
obtains [14] 
\begin{equation}
\Delta _{N}=\frac{\partial ^{2}}{\partial r^{2}}-\frac{\left( k-1\right)
\left( k-3\right) }{4r^{2}},
\end{equation}
and

\[
\Delta _{N}^{2}=\frac{\partial ^{4}}{\partial r^{4}}-\frac{\left( k-1\right)
\left( k-3\right) }{2r^{2}}\frac{\partial ^{2}}{\partial r^{2}}+\frac{\left(
k-1\right) \left( k-3\right) }{r^{3}}\frac{\partial }{\partial r} 
\]
\begin{equation}
+\frac{\left( k-1\right) ^{2}\left( k-3\right) ^{2}-24\left( k-1\right)
\left( k-3\right) }{16r^{4}}.
\end{equation}
Thus, using Eqs (7) and (8) and after a lengthy manipulation but
straightforward, we may write Eq.(5) in more simple and explicit form as

\begin{equation}
\left[ -\frac{1}{2\mu }\frac{d^{2}}{dr^{2}}+\frac{\left[ \overline{k}-\left(
1-a\right) \right] \left[ \overline{k}-\left( 3-a\right) \right] }{8\mu r^{2}%
}+W_{n,l}(r)-\frac{W_{n,l}(r)^{2}}{2m%
{\acute{}}%
}\right] u_{n,l}(r)=0,
\end{equation}
with 
\begin{equation}
W_{n,l}(r)=V(r)-E_{n,l},
\end{equation}
and the effective mass 
\begin{equation}
m^{^{\prime }}=\eta ^{3}/\mu ^{2}=(m_{q}m_{Q}\mu )/(m_{q}m_{Q}-3\mu ^{2}).
\end{equation}
It is worthwhile to note that the expression (10) is in complete agreement
with the expansion formula made by Durand and Durand in Ref. [27]. The
perturbation term, $W_{n,l}(r)^{2};$ that is, $(v^{2}/c^{2})$ term in Eq.
(10) is significant only when it is small (i.e., $W_{n,l}(r)/m^{^{\prime
}}\ll 1)$. This condition is verified by the confining potentials used to
describe heavy$-$quark systems except near the color$-$Coulomb singularity
at the origin, and for $r\rightarrow \infty $. However, it is always
satisfied on the average as stated by Durand and Durand [27]. We now proceed
to solve Eq. (10) by applying the SLNET, obtaining an exact result in series
for the energy [12-14]. This is a fundamental feature that allows one to
attack problems that do not involve small coupling constant or Hamiltonians
without a solvable strong term. Such an approximation has been used for the
solution of nonrelativistic as well as relativistic wave equations with
spherically symmetric potentials [12-14] yielding sufficiently accurate
results. Imbo {\it et al.} [12] showed the high accuracy and the
applicability of this method to a large number of spherically symmetric
potentials{\it .} In the SLNET [9] it is convenient to shift the origin of
coordinates to $r=r_{0}$ $($or $x=0)$ by defining

\begin{equation}
x=\bar{k}^{1/2}(r-r_{0})/r_{0}.
\end{equation}
Expansions about this point in powers of $x$ and $\overline{k}$ yield
[9,13,14]

\begin{equation}
\frac{1}{r^{2}}=\frac{1}{r_{0}^{2}}\stackrel{\infty }{%
\mathrel{\mathop{\sum }\limits_{j=0}}%
}(-x)^{j}\frac{(1+j)}{\overline{k}^{-j/2}},
\end{equation}

\begin{equation}
V(x(r_{0}))\;=\frac{1}{Q}\stackrel{\infty }{%
\mathrel{\mathop{\sum }\limits_{j=0}}%
}\left( \frac{d^{j}V(r_{0})}{dr_{0}^{j}}\right) \frac{\left( r_{0}x\right)
^{j}}{j!}\overline{k}^{(4-j)/2},
\end{equation}
and also 
\begin{equation}
E_{n,l}\;=\frac{1}{Q}\stackrel{\infty }{%
\mathrel{\mathop{\sum }\limits_{j=0}}%
}\overline{k}^{(2-j)}E_{j}.
\end{equation}
Further, by substituting Eqs. (14) through (16) into Eq. (10), one gets

\[
\left\{ -\frac{1}{4\mu }\frac{d^{2}}{dx^{2}}+\left[ \frac{\overline{k}}{%
16\mu }-\frac{(2-a)}{8\mu }+\frac{(1-a)(3-a)}{16\mu \overline{k}}\right]
\times \stackrel{\infty }{%
\mathrel{\mathop{\sum }\limits_{j=0}}%
}(-x)^{j}\frac{\left( 1+j\right) }{\overline{k}^{j/2}}\right. 
\]

\[
+\frac{r_{0}^{2}\;}{Q}\stackrel{\infty }{%
\mathrel{\mathop{\sum }\limits_{j=0}}%
}\left( \frac{d^{j}V(r_{0})}{dr_{0}^{j}}\right) \frac{\left( r_{0}x\right)
^{j}}{j!}\overline{k}^{(2-j)/2}-\frac{r_{0}^{2}\;\overline{k}}{m^{\prime }Q}%
\left[ \stackrel{\infty }{%
\mathrel{\mathop{\sum }\limits_{j=0}}%
}\left( \frac{d^{j}V(r_{0})}{dr_{0}^{j}}\right) \frac{\left( r_{0}x\right)
^{j}}{j!}\overline{k}^{-j/2}\right] ^{2} 
\]

\begin{equation}
+\left. \frac{2r_{0}^{2}\;}{m^{\prime }Q}\stackrel{\infty }{%
\mathrel{\mathop{\sum }\limits_{j=0}}%
}\overline{k}^{(1-j)}E_{j}\times \stackrel{\infty }{%
\mathrel{\mathop{\sum }\limits_{j=0}}%
}\left( \frac{d^{j}V(r_{0})}{dr_{0}^{j}}\right) \frac{\left( r_{0}x\right)
^{j}}{j!}\overline{k}^{-j/2}\right\} \chi _{n_{r}}(x)={\cal E}_{n_{r}}\chi
_{n_{r}}(x),\newline
\end{equation}
with the eigenvalues 
\[
{\cal E}_{n_{r}}=\frac{r_{0}^{2}}{Q}\left\{ \bar{k}\left( E_{0}+\frac{%
E_{0}^{2}}{m^{^{\prime }}}\right) +\left( E_{1}+\frac{2E_{0}E_{1}}{%
m^{^{\prime }}}\right) +\left( E_{2}+\frac{2E_{0}E_{2}}{m^{^{\prime }}}+%
\frac{E_{1}^{2}}{m^{^{\prime }}}\right) \frac{1}{\bar{k}}\right. 
\]

\begin{equation}
\;+\left. \left( E_{3}+\frac{2E_{0}E_{3}}{m^{^{\prime }}}+\frac{2E_{1}E_{2}}{%
m^{^{\prime }}}\right) \frac{1}{\bar{k}^{2}}+\cdots \right\} .~~~~~~~~~~~~~~
\end{equation}
The parameter $Q$ is an arbitrary scale, but is to be set equal to $%
\overline{k}^{2}$ to rescale potential after the problem is solved formally
in the $1/\bar{k}$ expansion. Thus comparing Eqs. (17) and (18) with its
counterpart Schr{\"{o}}dinger$-$like equation for the one$-$dimensional
anharmonic oscillator problem which has been investigated in detail for
spherically symmetric case by Imbo {\it et al.} [12] . The final analytic
expression in the $1/\bar{k}$ expansion of the energy eigenvalues
appropriate to the SS particle is

\[
{\cal E}_{n_{r}}=\bar{k}\left[ \frac{1}{16\mu }+\frac{r_{0}^{2}V(r_{0})}{Q}-%
\frac{r_{0}^{2}V(r_{0})^{2}}{m^{^{\prime }}Q}+\frac{2r_{0}^{2}E_{0}V(r_{0})}{%
m^{^{\prime }}Q}\right] 
\]

\[
+\left[ (1+2n_{r})\frac{\omega }{2}-\frac{(2-a)}{8\mu }\right] +\frac{1}{%
\bar{k}}\left[ \frac{2r_{0}^{2}E_{2}V(r_{0})}{m^{^{\prime }}Q}+\beta ^{(1)}%
\right] 
\]

\begin{equation}
+\frac{1}{\bar{k}^{2}}\left[ \frac{2r_{0}^{2}E_{3}V(r_{0})}{m^{^{\prime }}Q}%
+\beta ^{(2)}\right] +O\left[ \frac{1}{\bar{k}^{3}}\right] ,
\end{equation}
but $n_{r}$ is to be set equal to $0,1,2,\cdots .$ The quantities $\beta
^{(1)}$ and $\beta ^{(2)}$ appearing in the correction to the leading order
of the energy expression are defined and listed in the Appendix A.

Thus, comparing the terms of Eq. (18) with their counterparts in Eq. (19)
and equating terms of same order in $\overline{k}$ yields the leading energy
[14]

\begin{equation}
E_{0}=V(r_{0})+\frac{m^{^{\prime }}}{2}\left[ \sqrt{1+\frac{Q}{%
4r_{0}^{2}m_{\mu }m^{^{\prime }2}}}-1\right] ,
\end{equation}
where $m_{\mu }=\mu /m^{^{\prime }}$and $r_{0}$ is chosen to minimize the
leading energy, Eq. (20), that is, [9,13-14]

\begin{equation}
\frac{dE_{0}}{dr_{0}}=0{~~~;~~}\frac{d^{2}E_{0}}{dr_{0}^{2}}>0.
\end{equation}
Therefore, $r_{0}$ satisfies the following expression 
\begin{equation}
r_{0}^{3}V^{\prime }(r_{0}){\left( \frac{m^{^{\prime }2}}{4}+\frac{Q}{%
16r_{0}^{2}m_{\mu }}\right) }^{1/2}=\frac{Q}{16m_{\mu }}.
\end{equation}
Further, to solve for the shifting parameter $a$, the next contribution to
the energy eigenvalue is chosen to vanish, (i.e., $E_{1}=0),$(cf. e.g. Refs.
[9,12-14]), which gives

\begin{equation}
a=2-4\mu (1+2n_{r})\omega ,
\end{equation}
where $\omega $ is given by

\begin{equation}
\omega =\frac{1}{4\mu }{\left[ 3+r_{0}V^{\prime \prime }(r_{0})/V^{\prime
}(r_{0})-16r_{0}^{4}m_{\mu }V^{\prime }(r_{0})^{2}/Q\right] }^{1/2},
\end{equation}
and $Q$ in Eq. (22) can be rewritten in a more convenient form as 
\begin{equation}
Q=8m_{\mu }{\left[ \;r_{0}^{2}V^{\prime }(r_{0})\right] }^{2}(1+\lambda ),
\end{equation}
with

\begin{equation}
\lambda =\sqrt{1+\left( \frac{{m^{^{\prime }}}}{{r_{0}V^{\prime }(r_{0})}}%
\right) ^{2}\;}.
\end{equation}
Therefore, solving Eqs. (23) through (26) together with $Q=\overline{k}^{2}$%
, yields

\begin{equation}
1+2l+4\mu (2n_{r}+1)\omega =2r_{0}^{2}V^{\prime }(r_{0})\left( 2m_{\mu
}+2m_{\mu }\lambda \right) ^{1/2},
\end{equation}
which is an explicit equation in $r_{0}$. Once $r_{0}$ is determined via Eq.
(27), $E_{0}$ can be obtained via Eq. (20), $E_{2}$ and $E_{3\text{ }}$are
also obtained by solving Eqs. (18) and (19). Consequently, the general
expression for the quark binding energy is

\begin{equation}
E_{n,l}=E_{0}+\frac{1}{r_{0}^{2}\left( 1-\frac{2W_{n,l}(r_{0})}{m^{^{\prime
}}}\right) }\left\{ \beta ^{(1)}+\frac{\beta ^{(2)}}{\bar{k}}+O\left[ \frac{1%
}{\bar{k}^{2}}\right] \right\} ,
\end{equation}
which is an elegant algebraic expression that gives a rapidly convergent
binding energy value with high accuracy. Now, in the framework of our
semi-relativistic independent particle model the mass levels for an atomlike 
$q\overline{Q}$ meson where a light quark $q$ is moving around an almost
fixed heavy anti-quark $\overline{Q},$ the Salpeter bound-state mass can
also be easily obtained from the expression of quark binding energy (28) as 
\begin{equation}
M_{n,l}(q\overline{Q})=m_{q}+m_{Q}+2E_{n,l}.
\end{equation}
Here, of course, we do not consider the recoil effects of the heavy
anti-quark $\overline{Q}.$ Finally, making use of Eqs. (27) through (29),
one can resolve Eq. (10) for different types of central potentials taking
into account the spin dependent terms $V_{SD}$ as correction terms to the
static potential. We also consider a common parameter set for the upsilon
and charmonium mass spectra in a flavour-independent case$.\newline
.$

\section{SOME POTENTIAL MODELS\noindent}

\subsection{Static potentials}

The $c\overline{b}$ system that we investigate in the context of SS wave
equation is often considered as nonrelativistic system, and consequently our
treatment is based upon Salpeter equation with a Hamiltonian 
\begin{equation}
H=H_{0}+H_{1}+V_{SD},
\end{equation}
where $H_{0}=-\Delta _{N}/2\mu +V(r)$ denotes the Schr\"{o}dinger Hamiltonian%
$,$ $H_{1}=-\Delta _{N}^{2}/8\eta ^{3}$ denotes the perturbation term and $%
V_{SD}$ is the spin-dependent term [1,4,9,21] given by\footnote{%
To the moment, the only measured splitting of $nS$-levels is that of $\eta
_{c}$ and $J/\psi ,$ which allows us to evaluate the so-called (c.o.g.) data
using $\overline{M}_{\psi }({\rm 1S})=(3M_{J/\psi }+M\eta _{c})/4$ and also $%
\overline{M}({\rm nS})=M_{V}({\rm nS})-(M_{J/\psi }-M\eta _{c})/4n$ [15,21].}
\begin{equation}
\text{ \ }V_{SD}\longrightarrow V_{SS}=\frac{32\pi \alpha _{s}}{9m_{q}m_{Q}}%
\delta ^{3}({\bf r}){\bf s}_{1}.{\bf s}_{2}.
\end{equation}
Spin-independent relativistic corrections are not included. For the purpose
of making some preliminary estimates of the energies of the lowest two ${\rm %
S}$-states of the $B_{c}$ system, it is necessary to consider only the
spin-spin part of the spin-dependent potential since these are ${\rm S}$%
-states. Our solutions to the SS equation are generated numerically for the
central potentials. The effects of the spin-dependent parts are added as a
perturbation as improvement for splitting the lowest ${\rm S}$-state of the
spin-averaged energies. The potential parameters in this section are all
strictly flavor-independent. The potential parameters and the constituent
mass parameters are fitted to the low-lying energy levels of charmonium and
upsilon systems. So, like most authors (cf. e.g. [1,4,8,9,21]), we determine
the coupling constant $\alpha _{s}(m_{c})$ from the well measured
experimental charmonium hyperfine splitting of the ${\rm 1S}(c\overline{c})$
state value of $\Delta E_{{\rm HF}}=M_{J/\psi }-M_{\eta _{c}}=117\pm 2~{\rm %
MeV}$. The numerical value of $\alpha _{s}$ is dependent on the potential
form and found to be compatible to the other measurements [1,3-4,6-9,15].
The observed $M_{J/\psi }-M_{\eta _{c}}$ hyperfine splitting fixes $\alpha
_{s}$ for each potential. The perturbative part of such a quantity was
evaluated at the lowest order in $\alpha _{s}.$ Baldicchi and Prosperi [6]
used the standard running QCD coupling expression

\begin{equation}
\alpha _{s}({\bf Q})=\frac{4\pi }{\left( 11-\frac{2}{3}n_{f}\right) \ln
\left( \frac{{\bf Q}^{2}}{\Lambda ^{2}}\right) }.
\end{equation}
with $n_{f}=4$ and $\Lambda =200~{\rm MeV}$ cut at a maximum value $\alpha
_{s}(0)=0.35$ to get the right $M_{J/\psi }-M_{\eta _{c}}$ splitting and to
treat properly the infrared region [6]. Whereas Brambilla and Vairo [3] took
in their perturbative analysis $0.26\leq \alpha _{s}(\mu =2~{\rm GeV})\leq
0.30.$

We consider now a class of static potentials of general form 
\begin{equation}
V(r)=-Ar^{-\gamma }+Br^{\gamma }+C_{0};\text{ \ \ }0<\gamma \leq 1,\text{ \ }%
A,B>0
\end{equation}
had been previously proposed by Lichtenberg {\it et al.} [17] in which the
additive constant $C_{0}$ may be of either sign. It was necessary to add a
substantial flavor-independent constant to each potential for each
quark-antiquark system. The constant C$_{0}$ in Eq. (33) is determined from
the fit to the spin-averaged data (SAD) or center-of-gravity (c.o.g.) mass
of the ${\rm 1S}$ state (i.e., $\overline{M}({\rm 1S})=3068~{\rm MeV})$ from
the PDG [30]$.$ Since the model is spin independent and as the energies of
the singlet states of quarkonium families have not been measured, a
theoretical estimate of these unknown levels introduces uncertainty into the
calculated SAD [9,31]. The best parameter fittings to (c.o.g.) data are
obtained via the chi-square value or the least-square fit $\chi ^{2}=%
\mathrel{\mathop{\sum }\limits_{i}}%
\left[ \frac{M_{i}^{th}-M_{ii}^{\exp t}}{\Delta M_{i}}\right] ^{2},$ where $%
\Delta M_{i}$ refer to the errors in the experimental data. The static
quarkonium potentials are monotone nondecreasing, and concave functions
which satisfy the condition

\begin{equation}
V^{\prime }(r)>0\text{ \ and \ }V^{\prime \prime }(r)\leq 0.
\end{equation}
This class of potentials, in the generality (33), comprises the following
family of potentials:\ \ \ \ \ \ \ \ \ \ \ \ \ \ \ \ \ \ \ \ \ \ \ \ \ \ \ \
\ \ \ \ \ \ \ \ \ \ \ \ \ \ \ \ \ \ \ \ \ \ \ \ \ \ \ \ \ \ \ \ \ \ \ \ \ \
\ \ \ \ \ \ \ \ \ \ \ \ \ \ \ \ \ \ \ \ \ \ \ \ \ \ \ \ \ \ \ \ \ \ \ \ \ \
\ \ \ \ \ \ \ \ \ \ \ \ \ \ \ \ \ \ \ \ \ \ \ \ \ \ \ \ \ \ \ \ \ \ \ \ \ \
\ \ \ \ \ \ \ \ \ \ \ \ \ \ \ \ \ \ \ \ \ \ \ \ \ \ \ \ \ \ \ \ \ \ \ \ \ \
\ \ \ \ \ \ \ \ \ \ \ \ \ \ \ \ \ \ \ \ \ \ \ \ \ \ \ \ \ \ \ \ \ \ \ \ \ \
\ \ \ \ \ \ \ \ \ \ \ \ \ \ \ \ \ \ \ \ \ \ 

\subsubsection{\it Cornell potential}

The QCD-motivated Coulomb-plus-linear potential (Cornell potential) has the
form [24]

\begin{equation}
V_{C}(r)=-\frac{\alpha _{c}}{r}+\kappa r+C_{0},
\end{equation}

The main drawback of this potential is that the $c\overline{c}$ and $b%
\overline{b}$ states lie in an intermediate region of quark separation where
neither limiting forms of (34) should be valid. \ 

{\bf a.} Coulomb and linear potential of Eichten {\it et al. }[1,22,24]:
with $\alpha _{c}=0.52,~\kappa =0.1756~{\rm GeV}^{{\rm 2}},$\ $C_{0}=-0.8138~%
{\rm GeV},~m_{c}=1.840~{\rm GeV},$and\ \ \ \ \ \ \ \ \ \ \ \ \ \ \ \ \ \ \ \
\ \ \ \ \ \ \ \ \ \ \ \ \ \ \ \ \ \ \ \ \ \ \ \ \ \ \ \ \ \ $~$ $m_{b}=5.232~%
{\rm GeV}.~$We label this potential by Cornell 1.

{\bf b.} Coulomb and linear potential (35) of Hagiwara {\it et al. }[18,22]:
with $\alpha _{c}=0.47,~\kappa =0.19~{\rm GeV}^{{\rm 2}},~C_{0}=51~{\rm MeV}%
,~m_{c}=1.32~{\rm GeV}~$and $m_{b}=4.729~{\rm GeV}.$ We label this potential
by Cornell 2.

\subsubsection{\it Song-Lin potential}

This phenomenological potential was proposed by Song and Lin [16] 
\begin{equation}
V_{SL}(r)=-Ar^{-1/2}+Br^{1/2}+C_{0},
\end{equation}
with $A=0.923~{\rm GeV}^{{\rm 1/2}},~B=0.511~{\rm GeV}^{{\rm 3/2}%
},~C_{0}=-0.760~{\rm GeV},~m_{c}=1.820~{\rm GeV},~$ and $m_{b}=5.190~{\rm GeV%
}.$ The characteristic features of this potential may be traced in Ref.
[16]. \ 

\subsubsection{\it Turin potential}

Lichtenberg {\it et al.} [17] suggested such a potential which is an
intermediate between the Cornell and Song-Lin potentials. It has the simple
form 
\begin{equation}
V_{T}(r)=-Ar^{-3/4}+Br^{3/4}+C_{0}\text{,}
\end{equation}
with $A=0.620$~${\rm GeV}^{{\rm 1/4}},~B=0.304~{\rm GeV}^{{\rm 7/4}},$~$%
C_{0}=-0.783~{\rm GeV},~m_{c}=1.790~{\rm GeV}~$ and$~~m_{b}=5.166~{\rm GeV}%
.~ $The characteristic feature of this potential can be traced in Ref. [17].

\subsubsection{Power-law potential}

{\bf a. }Power-law\ potential of Martin [19,22]{\it : }The phenomenological
power-law potential

\begin{equation}
V_{M}(r)=-8.056~{\rm GeV}+(6.898~{\rm GeV})(r\times 1~{\rm GeV})^{0.1},
\end{equation}
is labeled as Martin's potential [19] with $m_{c}=1.80~{\rm GeV}~$and$%
~~m_{b}=5.162~{\rm GeV}.$ \ \ \ \ \ \ 

{\bf b. }Power-law\ potential of Rosner et al. [20.22]:

\begin{equation}
V_{R}(r)=-0.726~{\rm GeV}+(0.801~{\rm GeV})\left[ (r\times 1~{\rm GeV}%
)^{\gamma }-1\right] /\gamma ,
\end{equation}
with $\gamma =-0.12,~m_{c}=1.56~{\rm GeV}~$and$~~m_{b}=4.96~{\rm GeV}.$\ \ \
\ \ \ \ \ \ \ \ \ \ \ \ \ \ \ \ \ \ \ \ \ 

\subsubsection{\it Logarithmic potential of Quigg and Rosner [21,22]}

A Martin's power-law potential reduces [19] into 
\begin{equation}
V_{L}(r)=-0.6161~{\rm GeV}+(0.733~{\rm GeV})\ln (r\times 1\text{ }{\rm GeV}),
\end{equation}
with $m_{c}=1.50~{\rm GeV}~$and$~m_{b}=4.890~{\rm GeV}.$ The potential forms
in (38), and (40) were used by Eichten and Quigg [1,21]. Further, all of
these potential forms were also used for $\psi $ and $\Upsilon $ data
probing $0.1$ ${\rm fm}<r<1$ ${\rm fm}$\ region [15].

\subsection{QCD-motivated potentials\ \ \ \ \ \ \ \ \ \ \ \ \ \ \ \ \ \ \ \
\ \ \ \ \ \ \ \ \ \ \ \ \ \ \ \ \ \ \ \ \ \ \ \ \ \ \ \ \ \ \ \ \ \ \ \ }

\subsubsection{\it Igi-Ono potential}

The Cornell, logarithmic and power law potentials for some systems should
account properly for the running $\alpha _{s}$ at the different distances.
The $\alpha _{s}$ and the string constant $\kappa $ that describes the
long-distance behavior in (35) can vary independently. The $\Lambda _{QCD}$
used to evaluate $\alpha _{s}$ is related to a specific renormalization
scheme so that comparison with other calculations is possible. This leads to
expanding $\alpha _{s}$ to at least 2-loop order and use of nontrivial
methods in the interpolation between long-and short-distance behaviors. The
interquark potential at short distances has been calculated to 2-loop
calculations in the modified minimal-subtraction ($\overline{MS}$) scheme
[25]. Together with the 2-loop expression for $\alpha _{s},$ one has [10,25]

\begin{equation}
V(r)=-\frac{4\alpha _{s}(\mu )}{3r}+\left[ 1+\frac{\alpha _{s}(\mu )}{2\pi }%
(b_{0}ln\mu r+A)\right] ,
\end{equation}
and

\begin{equation}
\alpha _{s}(\mu )=\frac{4\pi }{b_{0}ln(\mu ^{2}/\Lambda _{\overline{MS}}^{2})%
}\left[ 1-\frac{b_{1}}{b_{0}^{2}}\frac{\ln ln(\mu ^{2}/\Lambda _{\overline{MS%
}}^{2})}{ln(\mu ^{2}/\Lambda _{\overline{MS}}^{2})}\right] ,
\end{equation}
with

\begin{equation}
\left[ b_{0,}~b_{1,}~A\right] =\left[ 11-\frac{2}{3}n_{f},~102-\frac{38}{3}%
n_{f},~b_{0}\gamma _{E}+\frac{31}{6}-\frac{5}{9}n_{f}\right] .
\end{equation}
Here $n_{f}$ is the number of flavors with mass below $\mu $ and $\gamma
_{E}=0.5772$ is the Euler's number. The renormalization scale $\mu $ is
usually chosen to be $1/r$ to obtain a simple form for $V(r)$. The
singularity in $\alpha _{s}(1/r)$ at $r=1/\Lambda _{\overline{MS}}=5$ ${\rm %
GeV}^{{\rm -1}}\approx 1$ ${\rm fm}$ for $\Lambda _{\overline{MS}}=200$ $%
{\rm MeV}.$ \ This singularity can be removed by the substitution

\begin{equation}
\ln \frac{1}{r^{2}\Lambda _{\overline{MS}}^{2}}\longrightarrow f(r)=\ln %
\left[ \frac{1}{r^{2}\Lambda _{\overline{MS}}^{2}}+b_{0}\right] .
\end{equation}
The constant $b_{0}$ is an adjustable parameter of the potential and will
not affect the perturbative part of the potential. Hence, setting $n_{f}=4$
and $n_{f}=5$ in Eq. (43), the one-gloun exchange part of the interquark
potential simply take the following two forms

\begin{equation}
V_{OGE}^{(n_{f}=4)}(r)=-\frac{16\pi }{25}\frac{1}{rf(r)}\left[ 1-\frac{462}{%
625}\frac{lnf(r)}{f(r)}+\frac{2\gamma _{E}+\frac{53}{75}}{f(r)}\right] ,
\end{equation}
and 
\begin{equation}
V_{OGE}^{(n_{f}=5)}(r)=-\frac{16\pi }{23}\frac{1}{rf(r)}\left[ 1-\frac{348}{%
529}\frac{lnf(r)}{f(r)}+\frac{2\gamma _{E}+\frac{43}{69}}{f(r)}\right] ,
\end{equation}
respectively, where $f(r)$ is the function given in Eq. (44). Further, the
long distance interquark potential grows linearly leading to confinement as 
\begin{equation}
V_{L}(r)=a_{0}r.
\end{equation}
Igi and Ono [25] proposed a potential whose general form

\begin{equation}
V^{(n_{f}=4,5)}(r)=V_{OGE}^{(n_{f}=4,5)}+a_{0}r+d_{0}re^{-g_{0}r}+C_{0},
\end{equation}
so as to interpolate smoothly between the two parts. They added
phenomenologically a term $d_{0}re^{-g_{0}r}$ to the potential so that to
adjust the intermediate range behavior by which the range of $\Lambda _{%
\overline{MS}}$ is extended keeping linearly rising confining potential.
Hence, the $\Lambda _{\overline{MS}}=(100,500)~{\rm MeV}$ keeping a pretty
good fit to the $c\overline{c}$ and $b\overline{b}$ data. \ \ \ \ \ \ \ \ \
\ \ \ \ \ \ \ \ \ \ \ \ \ \ \ \ \ \ \ \ \ \ \ \ \ \ \ \ \ \ \ \ \ \ \ \ \ \
\ \ \ \ \ 

\subsubsection{\it Improved Chen-Kuang potential}

Chen and Kuang [26] proposed two improved potential models so that the
parameters in Ref. [26] all vary explicitly with $\Lambda _{\overline{MS}}$
so that these parameters can only be given numerically for several values of 
$\Lambda _{\overline{MS}}.$ Such potentials have the natural QCD
interpretation and explicit $\Lambda _{\overline{MS}}$ dependence both for
giving clear link between QCD and experiment and for convenience in
practical calculation for a given value of $\Lambda _{\overline{MS}}$. This
potential takes the general form 
\begin{equation}
V^{(n_{f}=4)}(r)=-\frac{16\pi }{25}\frac{1}{rf(r)}\left[ 1-\frac{462}{625}%
\frac{lnf(r)}{f(r)}+\frac{2\gamma _{E}+\frac{53}{75}}{f(r)}\right] +\kappa r,
\end{equation}
where the string tension is related to Regge slope by $\kappa =\frac{1}{2\pi
\alpha 
{\acute{}}%
}$. The function~ $f(r)$ in Eq. (49), takes the general form

\begin{equation}
f(r)=ln\left[ \frac{1}{\Lambda _{\overline{MS}}r}+4.62-A(r)\right] ^{2},
\end{equation}
and

\begin{equation}
A(r)=\left[ 1-\frac{1}{4}\frac{\Lambda _{\overline{MS}}}{\Lambda _{\overline{%
MS}}^{I}}\right] \frac{1-\exp \left\{ -\left[ 15\left[ 3\frac{\Lambda _{%
\overline{MS}}^{I}}{\Lambda _{\overline{MS}}}-1\right] \Lambda _{\overline{MS%
}}r\right] ^{2}\right\} }{\Lambda _{\overline{MS}}r},
\end{equation}
with the following set of parameters fit

\begin{equation}
\left[ \kappa ,~\alpha 
{\acute{}}%
,~\Lambda _{\overline{MS}}^{I}\right] =\left[ 0.1491\text{ }{\rm GeV}^{{\rm 2%
}},~1.067\text{ }{\rm GeV}^{{\rm -2}},~180\text{ }{\rm MeV}\right] .
\end{equation}
together with $m_{c}=1.478~{\rm GeV},~m_{b}=4.876~{\rm GeV},~\Lambda _{%
\overline{MS}}=(100,600)~{\rm MeV}~$and $\alpha _{s}=0.235.$ The details of
this potential can be traced in Ref. [26].

\section{SUMMARY AND CONCLUSIONS}

We have solved numerically SS wave equation, Eq. (3), for various potentials
to determine the position of the spin-averaged ground-state masses of $%
\overline{M}_{{\rm 1S}}(J/\psi ~$and $\eta _{c})=3068~{\rm MeV},$ $\overline{%
M}_{{\rm 1S}}(\Upsilon $ and $\eta _{b})=9447~{\rm MeV},\ $and also $%
\overline{M}_{{\rm 1S}}(B_{c})$ with the help of Eq. (29)$.$ For simplicity
and for the sake of comparison, we have neglected the variation of $\alpha
_{s}$ with momentum in (32) to have a common spectra for all states and
scale the splitting of $c\overline{b}$ and $b\overline{b}$ from the
charmonium value. In our calculations for the hyperfine nsplitting values of
the $c\overline{b}$ spectrum, we have followed the same approach of Ref. [9]
in fitting the QCD coupling constant $\alpha _{s}.$

Table I reports the SS SAD mass spectrum and hyperfine splitting values of
the vector and pseudoscalar masses of the c$\overline{b}$ for a family of
seven central potentials. Masses and splittings lie within the ranges quoted
by Kwong and Rosner [7], Eichten and Quigg [1] and also Fulcher {\it et al.}
[4]. The value of the coupling constant $\alpha _{s},$ is obtained from the
proper splitting values of the $c\overline{c}$ and $b\overline{b}$
quarkonium systems [1,4,6-9].

It has been stated the possibility of producing $c\overline{b}$ mesons in $%
e^{+}e^{-}$ and hadron-hadron colliders [31-34]. We have used the coupling
constant in our analysis in the range $0.185\leq \alpha _{s}\leq 0.293$ for
all central potentials and $0.190\leq \alpha _{s}\leq 0.448$ for the
QCD-motivated potentials as\ shown in Tables I and IV. Here, we have
followed most authors in fixing the coupling constant $\alpha _{s}(m_{c})$
in reproducing all the spectra, [c.f. e.g. Refs. [1,4,8,9]), although slight
changes are made by other authors [15,35].\footnote{%
Kiselev {\it et al.} [15] have taken into account that $\Delta M_{\Upsilon }(%
{\rm 1S})=\frac{\alpha _{s}(\Upsilon )}{\alpha _{s}(\psi )}\Delta M_{\psi }(%
{\rm 1S})$ with $\alpha _{s}(\Upsilon )/\alpha _{s}(\psi )\simeq 3/4.$ On
the other hand, Motyka and Zalewiski [35] also found $\frac{\alpha
_{s}(m_{b}^{2})}{\alpha _{s}(m_{c}^{2})}\simeq 11/18.$}

Our results for SAD\ c$\overline{b}$ masses together with their lowest ${\rm %
S}$-state splittings for both Schr\"{o}dinger and Salpeter models are
presented in Table II and compared with the other estimates of Refs.
[4,24,32,36]. The values of the $\chi ^{2}$ fits are listed in Table II,
they show that the Salpeter results yield $\chi ^{2}=447$ in the present
model and give better agreement with experiment, PDG [30], than the
Schr\"{o}dinger results of $\chi ^{2}=836.$ Thus, the $\chi ^{2}$ of the
overall fit increases by about a factor 2. Hence, we conclude that extending
the contest between the Salpeter equation and Schr\"{o}dinger's [9] to the
charmonium and bottomonium systems provides additional evidence of an
experimental signature for the semi-relativistic kinetic energy correction
operator (second term) of Eq. (3). We have found that the measured values
for the SAD of charmonium and the upsilon system are in better agreement
with the Salpeter equation results than the Schr\"{o}dinger equation
results, (cf. e.g. Ref. [9]) and Table II of the present work, which is the
same conclusion reached by Jacobs {\it et al. }[37] and Fulcher {\it et al.}
[4]. Therefore, minimizing the quantity $\chi ^{2}$ is a stringent argument
that indicates the accuracy of the semirelativistic model over the previous
model [9]. Overmore, in Table II, we have considered two different fitted
sets of parameters for reproducing the $c\overline{b}$ masses. We noted that
the values of quark masses increase if we are allowed to add an additional
constant $C_{0}$ as given in the potential generality (33) and decrease if
we drop it out. Hence, the overall fit increases by nearly a factor 4 in
this trend.

Bambilla and Vairo [3] have calculated the maximum final result of $\left(
M_{B_{c}^{\ast }}\right) _{pert}=6326_{-9}^{+29}$ ${\rm MeV},$ the upper
limit corresponds to the choice of parameters $\Lambda _{\overline{MS}%
}^{n_{f}=3}=350~{\rm MeV}$ and $\mu =1.2~{\rm GeV},$ while the lower limit
to $\Lambda _{\overline{MS}}^{n_{f}=3}=250~{\rm MeV}$ and $\mu =2.0~{\rm GeV}
$ as the best approximation to their perturbative calculation. Our
predictions for the $c\overline{b}$ fine and hyperfine splittings together
with the ones estimated by other authors are listed in Table III. The SLNET
fine and hyperfine splitting estimations \ for a group potentials are all
fall in the range demonstrated by other authors. Larger discrepancies among
the various methods occur for the ground and excited states [6].

The fitted set of parameters and SAD mass spectrum and their splittings of
the Igi-Ono potential [25] labeled by type 1, 2 and 3, are listed in Table
IV, V and VI. It is clear that the overall study seems likely to be pretty
good and the reproduced fine and hyperfine splittings of the states are also
reasonable. Further, it is worthwhile to note that the quark masses $m_{c}$
and $m_{b}$ increase as the values of $\Lambda _{\overline{MS}}$ increase.
This is explained in the following way. The absolute value of the
short-range asymptotic behavior of the potential in Eqs. (45) and (46)
decreases with increasing $\Lambda _{\overline{MS}}.$ In order to reproduce
the SAD masses we need larger quark masses for larger values of $\Lambda _{%
\overline{MS}},$ (cf. e.g. Table III of Ref. [9]). Clearly, this is well
noted in the $m_{c}$ values.

We have also calculated the predicted $c\overline{b}$ spectrum obtained from
the Chen-Kuang potential in Table VII. We see that for states below the
threshold, the deviations of the predicted spectra from the experimental SAD
are within several ${\rm MeV}$. We find that $m_{c}$ and $m_{b}$ are
insensitive to the variation of $\Lambda _{\overline{MS}}$ for this
potential. The $\Lambda _{\overline{MS}}$ dependence of the Chen-Kuang
potential is given in Eq. (49). The potential found to be more sensetive
especially for the lowest states and found not sensitive for higher states.
The effect of $\Lambda _{\overline{MS}}$ is clearly on the Coulombic part of
Eq. (49).

In Table VIII, in order to get some idea of an error estimate (precision)
for our model, we have calculated the low-lying two ${\rm S}$-state masses
of the singlet and triplet $B_{c}$ system which are completely dependent on
the running coupling constant $\alpha _{s}$ values, (cf. e.g. Table I),
determined in charmonium and the upsilon systems. Using the largest and
smallest values of $\alpha _{s}$ in Table I to determine the errors, we have
calculated $M_{B_{c}}=6234_{-14}^{+10}$ ${\rm MeV},$ $M_{B_{c}^{\ast
}}=6310_{-6}^{+3}$ ${\rm MeV},$ and $\ 69$ ${\rm MeV}\leq \Delta _{{\rm 1S}%
}\leq 80$ ${\rm MeV.}$ Our hyperfine splitting value is in exellent
agreement with the other estimates made by other authors. Clearly, the
precision of the experiments [2] requires a very substantial improvement to
be sensitive to the bound-state mass differences between the various
calculations listed in Table VIII.

We study some properties of the $B_{c}$ system like the pseudoscalar decay
constant given by the Van Royen-Weisskopf formula modified for color [4,38],
that is,~ 
\begin{equation}
f_{B_{c}}=\sqrt{\frac{3}{\pi M_{B_{c}}}}\left| R_{{\rm 1S}}(0)\right| ,
\end{equation}
with the nonrelativistic radial wavefunction at the origin [21]

\begin{equation}
\left| R_{{\rm 1S}}(0)\right| =\sqrt{4\pi }\left| \psi _{{\rm 1S}}(0)\right|
.
\end{equation}
Moreover, in Eq. (53), the Salpeter bound-state mass of the low-lying $%
(n=1,l=0)$ pseudoscalar $B_{c}$-state can be found via

\begin{equation}
M_{B_{c}}(0^{-})=m_{c}+m_{b}+2E_{1,0}-3\Delta E_{{\rm HF}}/4,
\end{equation}
and also the vector $B_{c}^{\ast }$-state via

\begin{equation}
M_{B_{c}^{\ast }}(1^{-})=m_{c}+m_{b}+2E_{1,0}+\Delta E_{{\rm HF}}/4.
\end{equation}
whereas the square-mass difference can be simply found via

\begin{equation}
\Delta M^{2}=M_{B_{c}^{\ast }}^{2}(1^{-})-M_{B_{c}}^{2}(0^{-})=2\Delta E_{%
{\rm HF}}\left[ m_{c}+m_{b}+2E_{1,0}-\Delta E_{{\rm HF}}/4\right] ,.
\end{equation}
with

\begin{equation}
\Delta E_{{\rm HF}}=\frac{8\alpha _{s}(\mu )}{9m_{c}m_{b}}\left| R_{{\rm 1S}%
}(0)\right| ^{2}.
\end{equation}
Using our estimates of $M_{B_{c}}$ and $M_{B_{c}^{\ast }}$ in Table II, we
find~ decay constant for SS and for S models. They are listed in Table IX
together with the results of other authors [1,4,39-41]. All of these results
are also in reasonable agreement with the lattice results [5].

The empirical result obtained by Collins {\it et al}. [42] for potential
model wave functions at the origin, that is,

\begin{equation}
\left| R_{B_{c}}(0)\right| ^{2}\simeq \left| R_{J/\psi }(0)\right|
^{1.3}\left| R_{\Upsilon }(0)\right| ^{0.7},
\end{equation}
provides another touchstone for our numerical work. From charmonium and
upsilon calculations, we get wave function at the origin for SS and for S as
reported in Table IX which are in excellent agreement with the result of
Fulcher [4], in which $\left| R_{B_{c}}(0)\right| ^{2}\simeq 1.81~{\rm GeV}^{%
{\rm 3}}.$ Moreover, the empirical relationship for the ground state
hyperfine splittings, that is,

\begin{equation}
M_{B_{c}^{\ast }}-M_{B_{c}}\simeq 0.7\left( M_{J/\psi }-M_{\eta _{c}}\right)
^{0.65}(M_{\Upsilon }-M_{\eta _{b}})^{0.35},
\end{equation}
yields a splitting of $63~{\rm MeV},$ about 1\% and 5\% lower than our
estimations for S and SS equations, respectively, as shown in Table II.\
Both of these results are in reasonable agreement with the results of
Collins {\it et al}. [42], Fulcher [4] and FCY [4] and other authors [39-41]
.

\acknowledgments
(S. M. Ikhdair) gratefully acknowledges Dr. Suat G\"{u}nsel the founder
president of the Near East University, Prof. Dr. \c{S}enol Bekta\c{s}, and
Prof. Dr. Fakhreddin Mamedov the vice presidents of the NEU for their
continuous support and encouragement.\newpage

\appendix

\section{Some Useful Expressions}

Here we list the analytic expressions of $\beta ^{(1)},$ $\beta
^{(2)},\epsilon _{j},$ and $\delta _{j}$ for the spinless Salpeter equation
\ 

\begin{eqnarray}
\beta ^{(1)} &=&\frac{(1-a)(3-a)}{16\mu }+\left[ (1+2n_{r})~\bar{\varepsilon}%
_{2}+3(1+2n_{r}+2n_{r}^{2})\bar{\varepsilon}_{4}\right] \;  \nonumber \\
&-&\omega ^{-1}\left[ \bar{\varepsilon}_{1}^{2}+6(1+2n_{r})~\bar{\varepsilon}%
_{1}\bar{\varepsilon}_{3}+(11+30n_{r}+30n_{r}^{2})~\bar{\varepsilon}_{3}^{2}%
\right] ,
\end{eqnarray}
\begin{eqnarray}
\beta ^{(2)} &=&\left[ (1+2n_{r})\bar{\delta}_{2}+3(1+2n_{r}+2n_{r}^{2})\bar{%
\delta}_{4}~+5(3+8n_{r}+6n_{r}^{2}+4n_{r}^{3})\bar{\delta}_{6}~\right. 
\nonumber \\
&-&\omega ^{-1}(1+2n_{r})\bar{\varepsilon}_{2}^{2}+12(1+2n_{r}+2n_{r}^{2})%
\bar{\varepsilon}_{2}\bar{\varepsilon}_{4}+2~\bar{\varepsilon}_{1}\bar{\delta%
}_{1}  \nonumber \\
&+&2(21+59n_{r}+51n_{r}^{2}+34n_{r}^{3})\bar{\varepsilon}_{4}^{2}+6(1+2n_{r})%
\bar{\varepsilon}_{1}\bar{\delta}_{3}~  \nonumber \\
&+&30(1+2n_{r}+2n_{r}^{2})\bar{\varepsilon}_{1}\bar{\delta}%
_{5}+2(11+30n_{r}+30n_{r}^{2})\bar{\varepsilon}_{3}\bar{\delta}_{3} 
\nonumber \\
&&+\left. 10(13+40n_{r}+42n_{r}^{2}+28n_{r}^{3})\bar{\varepsilon}_{3}\bar{%
\delta}_{5}+6(1+2n_{r})\bar{\varepsilon}_{3}\bar{\delta}_{1}\right] 
\nonumber \\
\text{\ }~ &+&\omega ^{-2}\left[ 4~\bar{\varepsilon}_{1}^{2}~\bar{\varepsilon%
}_{2}+36(1+2n_{r})\bar{\varepsilon}_{1}\bar{\varepsilon}_{2}\bar{\varepsilon}%
_{3}+8(11+30n_{r}+30n_{r}^{2})\bar{\varepsilon}_{2}\bar{\varepsilon}%
_{3}^{2}~\right.  \nonumber \\
&+&24(1+2n_{r})\bar{\varepsilon}_{1}^{2}\bar{\varepsilon}%
_{4}+8(31+78n_{r}+78n_{r}^{2})\bar{\varepsilon}_{1}\bar{\varepsilon}_{3}\bar{%
\varepsilon}_{4}  \nonumber \\
&&+\left. 12~(57+189n_{r}+225n_{r}^{2}+150n_{r}^{3})\bar{\varepsilon}_{3}^{2}%
\bar{\varepsilon}_{4}\right]  \nonumber \\
&-&\omega ^{-3}\left[ ~8~\bar{\varepsilon}_{1}^{3}\bar{\varepsilon}%
_{3}+108(1+2n_{r})\bar{\varepsilon}_{1}^{2}\bar{\varepsilon}%
_{3}^{2}+48(11+30n_{r}+30n_{r}^{2})\bar{\varepsilon}_{1}\bar{\varepsilon}%
_{3}^{3}\right.  \nonumber \\
&+&\left. 30(31+109n_{r}+141n_{r}^{2}+94n_{r}^{3})\bar{\varepsilon}_{3}^{4}%
\right] ,~~~~~~~~~~~~
\end{eqnarray}
where 
\begin{equation}
\bar{\varepsilon _{i}}~=~\frac{\varepsilon _{i}}{(4\mu \omega )^{i/2}}%
,~~~~i=1,2,3,4.~~~~~
\end{equation}
and

\begin{equation}
\bar{\delta _{j}}=\frac{\delta _{j}}{(4\mu \omega )^{j/2}},~~~~j=1,2,3,4,5,6.
\end{equation}
\begin{equation}
\varepsilon _{1}=\frac{(2-a)}{4\mu }~,{~~~}\varepsilon _{2}=-\frac{3}{8\mu }%
~(2-a),
\end{equation}
\[
\varepsilon _{3}~=-\frac{1}{4\mu }~+~\frac{r_{0}^{5}}{6Q}~\left[ ~V^{\prime
\prime \prime }(r_{0})~-~\frac{2V(r_{0})V^{\prime \prime \prime }(r_{0})}{%
m^{^{\prime }}}~\right. 
\]

\begin{equation}
-\left. ~\frac{6V^{\prime }(r_{0})V^{\prime \prime }(r_{0})}{m^{^{\prime }}}%
~+~\frac{2V^{\prime \prime \prime }(r_{0})E_{0}}{m^{^{\prime }}}\right] ,
\end{equation}
\[
~\varepsilon _{4}=\frac{5}{16\mu }+~\frac{r_{0}^{6}~~}{24Q}\left[ V^{\prime
\prime \prime \prime }(r_{0})-\frac{2V(r_{0})V^{\prime \prime \prime \prime
}(r_{0})}{m^{^{\prime }}}~\right. 
\]

\begin{equation}
-\left. ~\frac{8V^{\prime }(r_{0})V^{\prime \prime \prime }(r_{0})}{%
m^{^{\prime }}}-\frac{6V^{\prime \prime }(r_{0})V^{\prime \prime }(r_{0})}{%
m^{^{\prime }}}~+~\frac{2V^{\prime \prime \prime \prime }(r_{0})E_{0}}{%
m^{^{\prime }}}\right] ,
\end{equation}
\begin{equation}
\delta _{1}~=-\frac{(1-a)(3-a)}{8\mu }~+~\frac{2r_{0}^{3}E_{2}V^{\prime
}(r_{0})}{m^{^{\prime }}Q},
\end{equation}

\begin{equation}
\delta _{2}~=\frac{3(1-a)(3-a)}{16\mu }~+~\frac{r_{0}^{4}E_{2}V^{\prime
\prime }(r_{0})}{~m^{^{\prime }}Q},
\end{equation}
\begin{equation}
\delta _{3}=\frac{(2-a)}{2\mu }~;{~~~}\delta _{4}=-\frac{5(2-a)}{8\mu },
\end{equation}
\[
\delta _{5}=-\frac{3}{8\mu }+\frac{r_{0}^{7}}{120Q}~\left[ ~V^{\prime \prime
\prime \prime \prime }(r_{0})~-~\frac{2V(r_{0})V^{\prime \prime \prime
\prime \prime }(r_{0})}{m^{^{\prime }}}\right. 
\]

\begin{equation}
-\left. \frac{10V^{\prime }(r_{0})V^{\prime \prime \prime \prime }(r_{0})}{%
m^{^{\prime }}}-\frac{20V^{\prime \prime }(r_{0})V^{\prime \prime \prime
}(r_{0})}{m^{^{\prime }}}~+~\frac{2V^{\prime \prime \prime \prime \prime
}(r_{0})E_{0}}{m^{^{\prime }}}\right] ,
\end{equation}
\[
\delta _{6}=\frac{7}{16\mu }+\frac{r_{0}^{8}}{720Q}~\left[ V^{\prime \prime
\prime \prime \prime \prime }(r_{0})~-~\frac{2V(r_{0})V^{\prime \prime
\prime \prime \prime \prime }(r_{0})}{m^{^{\prime }}}~-~\frac{12V^{\prime
}(r_{0})V^{\prime \prime \prime \prime \prime }(r_{0})}{m^{^{\prime }}}%
\right. ~ 
\]

\begin{equation}
-\left. \frac{20V^{\prime \prime \prime }(r_{0})V^{\prime \prime \prime
}(r_{0})}{m^{^{\prime }}}-~\frac{30V^{\prime \prime }(r_{0})V^{\prime \prime
\prime \prime }(r_{0})}{m^{^{\prime }}}~+~\frac{2V^{\prime \prime \prime
\prime \prime \prime }(r_{0})E_{0}}{m^{^{\prime }}}\right] ,
\end{equation}

$\bigskip $ \newpage

\bigskip \bigskip

\bigskip\ $\ \ \ \ \ \ \ \ \ \ \ \ \ 
\begin{array}{l}
\text{TABLE I. }c\overline{b}\text{ quarkonium masses and hyperfine
splittings }\Delta _{nS}{}^{\ast }\text{(in }{\rm MeV})\text{ } \\ 
\text{of the lowest two }{\rm S}\text{-states calculated for a group of
static potentials (GSP). } \\ 
\begin{tabular}{lllllllll}
\hline\hline
States & [6] & Cornell 1 & Cornell 2 & Song-Lin & Turin & Martin & Rosner & 
Logarithmic \\ 
\tableline$\alpha _{s}(m_{c})$ &  & $0.293$ & $0.258$ & $0.242$ & $0.263$ & $%
0.235$ & $0.185$ & $0.200$ \\ 
$\overline{M}(1S)$ &  & $6289$ & $6296$ & $6288$ & $6291$ & $6286$ & $6294$
& $6294$ \\ 
$M(1^{3}S_{1})$ & $6327$ & $6312$ & $6313$ & $6308$ & $6311$ & $6304$ & $%
6312 $ & $6311$ \\ 
$M(1^{1}S_{0})$ &  & $6220$ & $6244$ & $6230$ & $6231$ & $6232$ & $6238$ & $%
6241$ \\ 
$\Delta _{1S}$ & $77$ & $91.7$ & $69.1$ & $77.9$ & $79.6$ & $72.2$ & $73.7$
& $69.4$ \\ 
$\overline{M}(2S)$ &  & $6877$ & $6876$ & $6853$ & $6859$ & $6865$ & $6882$
& $6870$ \\ 
$M(2^{3}S_{1})$ & $6906$ & $6886$ & $6885$ & $6861$ & $6868$ & $6874$ & $%
6889 $ & $6878$ \\ 
$M(2^{1}S_{0})$ &  & $6850$ & $6848$ & $6827$ & $6832$ & $6836$ & $6859$ & $%
6846$ \\ 
$\Delta _{2S}$ & $42$ & $36.1$ & $37.0$ & $34.1$ & $35.8$ & $37.6$ & $29.7$
& $32.4$ \\ 
$\overline{M}(3S)$ &  & $7247$ & $7267$ & $7181$ & $7214$ & $7202$ & $7185$
& $7190$ \\ 
$\overline{M}(4S)$ &  & $7545$ & $7586$ & $7422$ & $7491$ & $7444$ & $7387$
& $7412$ \\ 
$\overline{M}(1P)$ & $6754$ & $6759$ & $6741$ & $6731$ & $6735$ & $6721$ & $%
6770$ & $6744$ \\ 
$\overline{M}(2P)$ & $7154$ & $7133$ & $7140$ & $7089$ & $7109$ & $7102$ & $%
7113$ & $7104$ \\ 
$\overline{M}(3P)$ &  & $7395$ & $7470$ & $7347$ & $7398$ & $7367$ & $7334$
& $7347$ \\ 
$\overline{M}(1D)$ & 7028 & $7018$ & $7014$ & $6998$ & $7002$ & $7003$ & $%
7043$ & $7021$ \\ 
$\overline{M}(2D)$ & 7367 & $7332$ & $7353$ & $7272$ & $7304$ & $7290$ & $%
7282$ & $7284$ \\ 
$\overline{M}(3D)$ &  & $7562$ & $7649$ & $7488$ & $7556$ & $7509$ & $7454$
& $7479$ \\ \hline\hline
\end{tabular}
\\ 
^{\ast }\Delta _{nS}=M(n^{3}S_{1})-M(n^{1}S_{0}).
\end{array}
$

\bigskip

\bigskip

\bigskip \bigskip

$
\begin{array}{l}
\text{TABLE II. }c\overline{b}\text{ masses and }\Delta _{nS}\text{ (in }%
{\rm MeV}\text{) of the using Coulomb-} \\ 
\text{plus-linear potential (CPLP), in the Schr\"{o}dinger and Salpeter
models.} \\ 
\begin{tabular}{lllllllll}
\hline\hline
States & CPLP$^{a}$ & CPLP$^{b}$ & CPLP$^{b}$ & [4]$^{a}$ & [4]$^{b}$ & [32]
& [36] & [24] \\ 
\tableline$\alpha _{s}(m_{c})$ & $0.276$ & $0.257$ & $0.322$ &  &  &  &  & 
\\ 
$\overline{M}(1S)$ & $6339$ & $6303$ & $6297$ & $6338$ & $6314$ & 6301 & 6317
& 6315 \\ 
$M(1^{3}S_{1})$ & $6355$ & $6319$ & $6317$ &  &  &  & 6332 &  \\ 
$M(1^{1}S_{0})$ & $6291$ & $6253$ & $6235$ &  &  &  & 6270 &  \\ 
$\Delta _{1S}$ & $63.7$ & $66.2$ & $82.2$ &  &  &  & 62 &  \\ 
$\overline{M}(2S)$ & $6930$ & $6885$ & $6852$ & $6918$ & $6872$ & $6893$ & $%
6870$ & $7009$ \\ 
$M(2^{3}S_{1})$ & $6940$ & $6894$ & $6862$ &  &  &  & 6881 &  \\ 
$M(2^{1}S_{0})$ & $6902$ & $6856$ & $6823$ &  &  &  & 6835 &  \\ 
$\Delta _{2S}$ & $38$ & $38.5$ & $39.6$ &  &  &  & 46 &  \\ 
$\overline{M}(3S)$ & $7352$ & $7285$ & $7220$ & $7336$ & $7270$ & $7237$ & 
7225 &  \\ 
$\overline{M}(4S)$ & $7707$ & $7615$ & $7520$ &  &  &  &  &  \\ 
$\overline{M}(1P)$ & $6756$ & $6741$ & $6725$ & $6755$ & $6745$ & $6728$ & 
& 6735 \\ 
$\overline{M}(2P)$ & $7195$ & $7152$ & $7100$ & 7193 & 7157 & 7122 &  &  \\ 
$\overline{M}(3P)$ & $7563$ & $7493$ & $7410$ &  &  &  &  &  \\ 
$\overline{M}(1D)$ & $7036$ & $7019$ & $6980$ &  &  &  &  & 7145 \\ 
$\overline{M}(2D)$ & $7418$ & $7371$ & $7299$ &  &  &  &  &  \\ 
$\overline{M}(3D)$ & $7755$ & $7677$ & $7577$ &  &  &  &  &  \\ 
Parameters &  &  &  &  &  &  &  &  \\ 
$\alpha _{c}$ & $0.472$ & $0.437$ & $0.457$ &  &  &  &  &  \\ 
$\kappa (GeV^{2})$ & $0.191~$ & $0.203$ & $0.182$ &  &  &  &  &  \\ 
$C_{0}(GeV)$ & 0.0 & $0.0$ & $-0.790$ &  &  &  &  &  \\ 
$m_{c}(GeV)$ & $1.3205$ & $1.321$ & $1.79$ &  &  &  &  &  \\ 
$m_{b}(GeV)$ & $4.7485~$ & $4.731$ & $5.17$ &  &  &  &  &  \\ 
$\chi ^{2}$ & $836^{c}$ & $447^{c}$ & $2011^{c}$ &  &  &  &  &  \\ 
\hline\hline
\end{tabular}
\\ 
^{a}\text{In the Schr\"{o}dinger model,}^{b}\text{In the Salpeter model,}^{c}%
\text{Fitting to }c\overline{c}\text{ and }b\overline{b}\text{ spectra.}
\end{array}
$\bigskip

\bigskip \bigskip \bigskip $
\begin{array}{l}
\text{TABLE III. Fine and hyperfine splittings of }c\overline{b}\text{
states (in }{\rm MeV}\text{) using } \\ 
\text{GSP, IOP and CKP in our work compared with the other authors. } \\ 
\begin{tabular}{llllllll}
\hline\hline
States & Latt.[6] & Qua.[6] & Lin.[6] & [4] & IOP$^{\ast }$ & GSP$^{\ast }$
& CKP$^{\ast }$ \\ \hline
Fine Splitting &  &  &  &  &  &  &  \\ 
$M(2S)-M(1S)$ & $672\pm 120$ & $558$ & $533$ & $579$ & $577\pm 25$ & $%
578_{-13}^{+10}$ & $494_{-34}^{+29}$ \\ 
$M(3S)-M(1S)$ &  & $931$ & $899$ &  & $916_{-13}^{+20}$ & $921_{-30}^{+50}$
& $866_{-43}^{+45}$ \\ 
$M(4S)-M(1S)$ &  &  &  &  & $1176_{-31}^{+29}$ & $1178_{-85}^{+112}$ & $%
1149_{-31}^{+23}$ \\ 
$M(2P)-M(1P)$ &  & $382$ & $376$ &  & $375\pm 8$ & $370_{-27}^{+29}$ & $%
389\pm 1$ \\ 
$M(3P)-M(1P)$ &  &  &  &  & $657_{-10}^{+32}$ & $637_{-73}^{+92}$ & $%
695_{-1}^{+3}$ \\ 
$M(2D)-M(1D)$ &  & $324$ & $321$ &  & $296_{-12}^{+21}$ & $288_{-49}^{+51}$
& $318_{-1}^{+2}$ \\ 
$M(3D)-M(1D)$ &  &  &  &  & $547_{-18}^{+42}$ & $514_{-103}^{+121}$ & $%
586\pm 1$ \\ 
Hyperfine Splitting &  &  &  &  &  &  &  \\ 
$\Delta _{1S}$ & $41\pm 20$ & $77$ & $62$ & $55$ & $63\pm 1$ & $%
76_{-7}^{+16} $ & $65_{-13}^{+5}$ \\ 
$\Delta _{2S}$ & $30\pm 8$ & $42$ & $33$ & $32$ & $32$ & $35_{-5}^{+3}$ & $%
35_{-1.2}^{+0.2}$ \\ \hline
\end{tabular}
\\ 
\begin{array}{l}
^{\ast }\text{ The averaged splittings in our work for different potentials. 
}
\end{array}
\end{array}
$

\bigskip

\bigskip

\bigskip $
\begin{array}{l}
\text{TABLE IV. Parameters used for Igi-Ono potential (IOP).} \\ 
\begin{tabular}{llllll}
\hline\hline
$a_{0}~({\rm GeV}^{{\rm 2}})$ & $g_{0}~({\rm GeV})$ & $d_{0}~({\rm GeV}^{%
{\rm 2}})$ & $\alpha _{s}(m_{c})$ & $m_{c}~({\rm GeV})$ & $m_{b}~({\rm GeV})$
\\ \hline
Type 1 ($n_{f}=5)$ &  &  &  &  &  \\ 
0.1587 & $0.3436$ & 0.2550 & $0.190$ & 1.38 & 4.782 \\ \hline
Type 2 ($n_{f}=4)$ &  &  &  &  &  \\ 
0.1587 & 0.3436 & 0.2550 & $0.403$ & 1.353 & 4.746 \\ 
\tableline Type 3 ($n_{f}=4)$ &  &  &  &  &  \\ 
$0.1585$ & 0 & 0 & $0.448$ & $\text{1.494~}$ & $4.885~$ \\ \hline\hline
\end{tabular}
\end{array}
$

\bigskip

\bigskip

\bigskip \bigskip \bigskip

\bigskip $
\begin{array}{l}
\text{TABLE V. }c\overline{b}\text{ quarkonium mass spectrum and }\Delta
_{nS}\text{ (in }{\rm MeV}\text{)} \\ 
\text{predicted by using IOP of types 1, 2, and 3 in our work. } \\ 
\begin{tabular}{llllllll}
\hline\hline
States & $\Lambda _{\overline{MS}}({\rm MeV})=$ & 200 & 200 & 200 & 250 & 300
& 400 \\ 
$b$ &  & 20$^{a}$ & 20$^{b}$ & 23.3$^{c}$ & 23.3$^{c}$ & 23.3$^{c}$ & 23.3$%
^{c}$ \\ \hline
$\overline{M}(1S)$ &  & 6295 & 6294 & 6364 & 6324 & 6292 & 6245 \\ 
$M(1^{3}S_{1})$ &  & 6311 & 6310 & 6377 & 6340 & 6310 & 6267 \\ 
$M(1^{1}S_{0})$ &  & 6247 & 6247 & 6324 & 6277 & 6239 & 6179 \\ 
$\Delta _{1S}$ &  & 64.6 & 63.5 & 53.3 & 62.1 & 70.9 & 87.9 \\ 
$\overline{M}(2S)$ &  & 6897 & 6872 & 6899 & 6876 & 6857 & 6829 \\ 
$M(2^{3}S_{1})$ &  & 6905 & 6880 & 6907 & 6884 & 6865 & 6837 \\ 
$M(2^{1}S_{0})$ &  & 6873 & 6848 & 6876 & 6852 & 6832 & 6803 \\ 
$\Delta _{2S}$ &  & 32.2 & 32.2 & 31.2 & 32.3 & 33.1 & 34 \\ 
$\overline{M}(3S)$ &  & 7231 & 7197 & 7249 & 7233 & 7222 & 7210 \\ 
$\overline{M}(4S)$ &  & 7474 & 7439 & 7539 & 7529 & 7523 & 7516 \\ 
$\overline{M}(1P)$ &  & 6757 & 6733 & 6755 & 6741 & 6731 & 6719 \\ 
$\overline{M}(2P)$ &  & 7132 & 7099 & 7134 & 7124 & 7117 & 7108 \\ 
$\overline{M}(3P)$ &  & 7404 & 7368 & 7437 & 7430 & 7425 & 7420 \\ 
$\overline{M}(1D)$ &  & 7041 & 7008 & 7019 & 7013 & 7009 & 7005 \\ 
$\overline{M}(2D)$ &  & 7328 & 7292 & 7335 & 7330 & 7327 & 7323 \\ 
$\overline{M}(3D)$ &  & 7570 & 7532 & 7606 & 7602 & 7599 & 7597 \\ \hline
\end{tabular}
\\ 
^{a}\text{ Type 1 with }C_{0}=-28~{\rm MeV}\text{~to fit (c.o.g.) value.} \\ 
^{b}\text{ Type 2 with }C_{0}=-28~{\rm MeV}~\text{to fit (c.o.g.) value} \\ 
^{c}\text{ Type 3 with }C_{0}=-29~{\rm MeV}~\text{to fit (c.o.g.) value.}
\end{array}
$

\bigskip

$\bigskip 
\begin{array}{l}
\text{TABLE VI. }c\overline{b}\text{ quarkonium mass spectrum } \\ 
\text{and }\Delta M_{n}^{a}\text{ (in }{\rm MeV}\text{) predicted by using
IOP.}^{b}\text{ } \\ 
\begin{tabular}{llllllll}
\hline\hline
States & $\Lambda _{\overline{MS}}(MeV)$ & $200$ &  & $250$ &  & $300$ &  \\ 
\tableline$M({\rm 1S})$ &  & $6382$ &  & $6342$ &  & $6311$ &  \\ \hline
$M({\rm 2S})$ &  & $6918$ &  & $6894$ &  & $6875$ &  \\ 
$M({\rm 3S})$ &  & $7267$ &  & $7251$ &  & $7241$ &  \\ 
$M({\rm 4S})$ &  & $7558$ &  & $7548$ &  & $7542$ &  \\ 
$M({\rm 1P})$ &  & $6773$ &  & $6759$ &  & $6750$ &  \\ 
$M({\rm 2P})$ &  & $7153$ &  & $7142$ &  & $7135$ &  \\ 
$M({\rm 3P})$ &  & $7456$ &  & $7448$ &  & $7443$ &  \\ 
$M({\rm 1D})$ &  & $7037$ &  & $7031$ &  & $7027$ &  \\ 
$M({\rm 2D})$ &  & $7353$ &  & $7348$ &  & $7345$ &  \\ 
$M({\rm 3D})$ &  & $7625$ &  & $7620$ &  & $7618$ &  \\ 
$\Delta M_{{\rm 2S}}$ & $585^{c}$ & $536$ &  & $552$ &  & $564$ &  \\ 
$\Delta M_{{\rm 3S}}$ &  & $349$ &  & $357$ &  & $366$ &  \\ 
$\Delta M_{{\rm 4S}}$ &  & $291$ &  & $297$ &  & $301$ &  \\ 
$\Delta M_{{\rm 2P}}$ &  & $380$ &  & $383$ &  & $385$ &  \\ 
$\Delta M_{{\rm 3P}}$ &  & $303$ &  & $306$ &  & $308$ &  \\ 
$\Delta M_{{\rm 2D}}$ &  & $316$ &  & $317$ &  & $318$ &  \\ 
$\Delta M_{{\rm 3D}}$ &  & $272$ &  & $272$ &  & $273$ &  \\ \hline
\end{tabular}
\\ 
^{a}\text{Fine splitting is }\Delta M_{(n+1){\rm S}}=M((n+1){\rm S})-M(n{\rm %
S}). \\ 
^{b}\text{Type 3 fitted to the experimental }c\overline{c}\text{ and }b%
\overline{b}~\text{spectra with }m_{b}=4.874~{\rm GeV}. \\ 
^{c}\text{Here we cite GKLT in Ref. [32].}
\end{array}
$

\bigskip

$\bigskip $\bigskip

$
\begin{array}{l}
\text{TABLE VII. }c\overline{b}\text{ quarkonium mass spectrum and }\Delta
_{nS}\text{ } \\ 
\text{(in }{\rm MeV}\text{) predicted by using CKP in our work.} \\ 
\begin{tabular}{llllllll}
\hline\hline
States & $\Lambda _{\overline{MS}}({\rm MeV})$ & $100-270$ & $275-350$ & $%
400 $ & $450$ & $500$ & $600$ \\ \hline
$\overline{M}({\rm 1S})$ &  & $6250$ & $6308$ & $6348$ & $6278$ & $6318$ & $%
6277$ \\ 
$M({\rm 1}^{{\rm 3}}{\rm S}_{{\rm 1}})$ &  & $6267$ & $6325$ & $6365$ & $%
6293 $ & $6331$ & $6293$ \\ 
$M({\rm 1}^{{\rm 1}}{\rm S}_{{\rm 0}})$ &  & $6197$ & $6255$ & $6296$ & $%
6232 $ & $6279$ & $6231$ \\ 
$\Delta _{{\rm 1S}}$ &  & $69.9$ & $69.9$ & $68.9$ & $68.9$ & $51.8$ & $61.6$
\\ 
$\overline{M}({\rm 2S})$ &  & $6808$ & $6808$ & $6808$ & $6808$ & $6841$ & $%
6835$ \\ 
$M({\rm 2}^{{\rm 3}}{\rm S}_{{\rm 1}})$ &  & $6817$ & $6817$ & $6817$ & $%
6817 $ & $6850$ & $6844$ \\ 
$M({\rm 2}^{{\rm 1}}{\rm S}_{{\rm 0}})$ &  & $6782$ & $6782$ & $6782$ & $%
6782 $ & $6816$ & $6809$ \\ 
$\Delta _{{\rm 2S}}$ &  & $35.2$ & $35.2$ & $35.2$ & $35.2$ & $33.8$ & $35.1$
\\ 
$\overline{M}({\rm 3S})$ &  & $7171$ & $7171$ & $7171$ & $7171$ & $7229$ & $%
7173$ \\ 
$\overline{M}({\rm 4S})$ &  & $7466$ & $7466$ & $7466$ & $7466$ & $7490$ & $%
7466$ \\ 
$\overline{M}({\rm 1P})$ &  & $6674$ & $6674$ & $6674$ & $6674$ & $6690$ & $%
6678$ \\ 
$\overline{M}({\rm 2P})$ &  & $7062$ & $7062$ & $7062$ & $7062$ & $7080$ & $%
7065$ \\ 
$\overline{M}({\rm 3P})$ &  & $7368$ & $7368$ & $7368$ & $7368$ & $7388$ & $%
7368$ \\ 
$\overline{M}({\rm 1D})$ &  & $6953$ & $6953$ & $6953$ & $6953$ & $6957$ & $%
6955$ \\ 
$\overline{M}({\rm 2D})$ &  & $7270$ & $7270$ & $7270$ & $7270$ & $7277$ & $%
7270$ \\ 
$\overline{M}({\rm 3D})$ &  & $7540$ & $7540$ & $7540$ & $7540$ & $7542$ & $%
7540$ \\ \hline
\end{tabular}
\\ 
^{a}\text{ Type 1 with }C_{0}=-28~{\rm MeV}~\text{~to agree with (c.o.g.)
value.}
\end{array}
$

\bigskip $
\begin{array}{l}
\text{TABLE VIII. The predicted }c\overline{b}\text{ masses of the lowest }%
{\rm S}\text{-wave and its } \\ 
\text{splitting (all in }{\rm MeV})\text{ compared with the other authors. }
\\ 
\bigskip 
\begin{tabular}{lllllll}
\hline\hline
work & $\overline{M}({\rm 1S}$) & $M_{B_{c}}(1^{1}S_{0})^{\ast }$ &  & $%
M_{B_{c}^{\ast }}(1^{3}S_{1})$ & $\Delta _{{\rm 1S}}$ &  \\ 
\tableline Eichten {\it et al.} [1] &  & $6258\pm 20$ &  &  &  &  \\ 
Colangelo {\it et al.} [3] &  & $6280$ &  & $6350$ & $70$ &  \\ 
Baker {\it et al. }[33] &  & $6287$ &  & $6372$ & $85$ &  \\ 
Roncaglia {\it et al.} [33] &  &  &  & $6320\pm 10$ &  &  \\ 
Godfrey {\it et al.} [1] &  & $6270$ &  & $6340$ &  &  \\ 
Bagan {\it et al.} [1,33] &  & $6255\pm 20$ &  & $6330\pm 20$ & 75 &  \\ 
Bambilla{\it \ et al.} [3] &  &  &  & $6326_{-9}^{+29}$ &  &  \\ 
Baldicchi {\it et al. }[6] &  & $6194\sim 6292$ &  & $6284\sim 6357$ & $%
65\leq \Delta _{{\rm 1S}}\leq 90$ &  \\ 
SLNET (GSP)$^{\dagger }$ & $6291\pm 5$ & $6234_{-14}^{+10}$ &  & $%
6310_{-6}^{+3}$ & $69\leq \Delta _{{\rm 1S}}\leq 80$ &  \\ 
SLNET (IOP)$^{\ddagger }$ & $6324$ & $6277$ &  & $6340$ & $62$ &  \\ 
SLNET & $6291^{\intercal }$ & $6234$ &  & $6310$ & $66$ &  \\ \hline
\end{tabular}
\\ 
^{\ast }\text{ Experimental mass of such a singlet state is presented in
[2,6].} \\ 
^{\dagger }\text{ Averaging over the seven values in Table I.} \\ 
^{\ddagger }\text{ An estimation using type 3 with }\Lambda _{\overline{MS}%
}=250~{\rm MeV}. \\ 
^{\intercal }\text{ Best estimation to the (c.o.g.) lowest }{\rm S}\text{%
-state.}
\end{array}
$

\bigskip

\bigskip

\bigskip

\begin{center}
$
\begin{array}{l}
\text{TABLE IX. The leptonic }B_{c}\text{ meson constant and radial wave
function } \\ 
\text{at the origin calculated in our model and by the other authors.} \\ 
\begin{tabular}{llllllllll}
\hline\hline
Level ($1^{1}S_{0}$) & Martin & Coulomb & EQ[1] & F[4] & [39] & [40] & [41]
& S$^{\ast }$ & SS$^{\ast }$ \\ 
\tableline$f_{B_{c}}~({\rm MeV})$ & $510\pm 80$ & $456\pm 70$ & 495 & 517 & 
410 & 600 & 500 & 495 & 524 \\ 
$\left| R_{B_{c}}(0)\right| ^{2}~({\rm GeV}^{{\rm 3}})$ & $1.716$ & - & $%
1.638$ & 1.81 & - & - & - & 1.85 & 1.90 \\ \hline\hline
\end{tabular}
\\ 
^{\ast }\text{Present work.}
\end{array}
$\bigskip
\end{center}

\bigskip

\bigskip

\bigskip

\bigskip

\bigskip

\bigskip

\bigskip

\bigskip

\bigskip

\bigskip

\bigskip

\end{document}